%% file: paper.tex
\documentclass[twocolumn,showpacs,aps,prl,superscriptaddress]{revtex4}
\usepackage{graphicx}
\usepackage{dcolumn}
\usepackage{epsfig}
\usepackage{amsmath}

\RequirePackage{xspace}

\newcommand{\BaBarYear}    {07}
\newcommand{\BaBarNumber}  {056}
\newcommand{\SLACPubNumber} {12826}
\def\epem  {\ensuremath{e^+e^-}\xspace}

\newcommand{\thetaT}{\ensuremath{\theta_{\rm T}}}
\newcommand{\costhr}{\ensuremath{\cos\thetaT}}
\newcommand{\hel}{\ensuremath{{\cal H}}}

\def\ra                 {\ensuremath{\rightarrow}\xspace}
\newcommand{\xf}{\mbox{${\cal F}$}}
\def\KS    {\ensuremath{K^0_{\scriptscriptstyle S}}}
\def\Bu      {\ensuremath{B^+}}

\def\Bub     {\ensuremath{B^-}}
\def\Bbar    {\overline{B}{}}

\def\qq{\mbox{$q\bar q\ $}}
\def\Bzb     {\ensuremath{\Bbar^0}}
\def\Bz      {\ensuremath{B^0}}
\def\BpBm    {\ensuremath{\Bu {\kern -0.16em \Bub}}\xspace}
\def\BzBzb   {\ensuremath{\Bz {\kern -0.16em \Bzb}}\xspace}
\def\CP                {\ensuremath{C\!P}\xspace}

\newcommand{\DE}{\ensuremath{\Delta E}}
\newcommand{\pvec}{{\bf p}}
\newcommand{\half}{\mbox{${1\over2}$}}
 \def\mes{\mbox{$m_{\rm ES}$}}
\newcommand{\calB}{\mbox{${\cal B}$}}
\newcommand{\auno}{\mbox{$a_1(1260)$}}

\newcommand{\aduek} {\mbox{$a^-_2(1320)\, K^+$}}

\newcommand{\Nbtoappim}{\mbox{$B^0 \rightarrow a^{\pm}_1(1260)\, \pi^{\mp}  $}}

\newcommand{\Nbntoapk}{\mbox{$B^0 \rightarrow a^{-}_1(1260)\,  K^{+}  $}}
\newcommand{\Nbctoapk}{\mbox{$B^{+} \rightarrow a^{+}_1(1260)\,  K^{0}  $}}

\newcommand{\Nbnapkmm}{\mbox{$a^{-}_1(1260)\,  K^{+}  $}}
\newcommand{\Nbcapkm}{\mbox{$a^{+}_1(1260)\,  K^{0}  $}}

\newcommand{\Nbntoaduek}{\mbox{$B^0 \rightarrow a^{-}_2(1320)\, K^{+}  $}}
\newcommand{\Nbctoaduek}{\mbox{$B^+ \rightarrow a^{+}_2(1320)\, K^{0}$}}
\newcommand{\NbKpi}{\mbox{$B \rightarrow K_1(1270)\,\pi$}}
\newcommand{\NbKppi}{\mbox{$B \rightarrow K_1(1400)\,\pi$}}

\newcommand{\NNatorhopi}{\mbox{$ a_1(1260) \rightarrow \rho \pi  $}}

\newcommand{\Natopipipi}{\mbox{$ a^{\pm}_1(1260) \rightarrow \pi^{\mp}\pi^{\pm}\pi^{\pm}  $}}
\newcommand{\Naptopipipi}{\mbox{$ a^{+}_1(1260) \rightarrow \pi^{-}\pi^{+}\pi^{+}  $}}
\newcommand{\Namtopipipi}{\mbox{$ a^{-}_1(1260) \rightarrow \pi^{+}\pi^{-}\pi^{-}  $}}

\newcommand{\UfourS}{\mbox{$\Upsilon(4S)$}}

\newcommand{\BrakcPiPiPi}{\mbox{$\calB(\Nbctoapk) \, \calB(\Naptopipipi)   $}}
\newcommand{\BraknPiPiPi}{\mbox{$\calB(\Nbntoapk)  \,  \calB(\Namtopipipi)    $}}

\newcommand{\rakn}{\mbox{$8.2 \pm 1.5 \pm 1.2 $}}
\newcommand{\Rakn}{\mbox{$(\rakn)\times 10^{-6}$}}
\newcommand{\rakc}{\mbox{$17.4 \pm 2.5 \pm 2.2 $}}
\newcommand{\Rakc}{\mbox{$(\rakc)\times 10^{-6}$}}

\def\BB{\mbox{$B\overline B\ $}}
\def\pep2{PEP-II}

\newcommand\etal{{\it et al.}}
\newcommand{\dedx}{\ensuremath{\mathrm{d}\hspace{-0.1em}E/\mathrm{d}x}}
\newcommand{\gevcc}{\mbox{$\textrm{GeV}/c^2$}} 
\newcommand{\mevcc}{\mbox{$\textrm{MeV}/c^2$}} 
\newcommand{\gevc}{\mbox{$\textrm{GeV}/c$}} 
\newcommand{\gev}{\mbox{$\textrm{GeV}$}} 
\newcommand{\mev}{\mbox{$\textrm{MeV}$}} 
 
\newcommand{\jprlBase}  [1]     {Phys.\ Rev.\ Lett.}
\newcommand{\jprl}      [1]    {\jprlBase\ ~{\bf #1}}
\newcommand{\jprBase}        {Phys.\ Rev.\ } 
\newcommand{\jprd}      [1]  {\jprBase\ D~{\bf #1}}
\newcommand{\plBase}   [1]         {Phys.\ Lett.}
\newcommand{\plb}      [1]    {\plBase\ B~{\bf #1}}
\newcommand{\nimBaseA}       {Nucl.\ Instr.\ Methods Phys.\ Res., Sect.\ A\xspace}
\newcommand{\nima}      [1]  {\nimBaseA~{\bf #1}}
\newcommand{\zpBase}         {Z.\ Phys.}
\newcommand{\zpc}       [1]  {\zpBase\ C~{\bf #1}}

\newcommand{\jpg}       [1]  {{J.\ Phys.\ {\bf G{\bf #1}}}}
\newcommand{\progtp}    [1]  {{Prog.\ Theor.\ Phys.\ {\bf #1}}}
\newcommand{\epjBase}  [1]     {Eur.\ Phys.\ J. \xspace}
\newcommand{\epj}      [1]    {\epjBase\ C~{\bf #1}}

\usepackage{relsize}
\def\babar{\mbox{\slshape B\kern-0.1em{\smaller A}\kern-0.1em
    B\kern-0.1em{\smaller A\kern-0.2em R}}}

\def\figurebox#1#2#3{%
    \def\arg{#3}%
    \ifx\arg\empty
    {\hfill\vbox{\hsize#2\hrule\hbox to #2{\vrule\hfill\vbox to #1{\hsize#2\vfill}\vrule}\hrule}\hfill}%
    \else
    {\hfill\epsfbox{#3}\hfill}%
    \fi}

\begin{document}
 
\begin{flushleft}
\babar-PUB-\BaBarYear/\BaBarNumber \\
SLAC-PUB-\SLACPubNumber\\
\end{flushleft}

\title{\large  \bf\boldmath  Observation of \Nbctoapk\ and \Nbntoapk\ }
 
\input authors_jul2007.tex

\begin{abstract}
\noindent
We present  branching fraction measurements of the decays \Nbctoapk\ and \Nbntoapk\ with \Natopipipi. 
The data sample corresponds to $383\times 10^6$ \BB\ pairs  produced in \epem\ annihilation through the
\UfourS\ resonance. We measure the products of the branching fractions \BrakcPiPiPi =\Rakc\ and \BraknPiPiPi = \Rakn. We
also measure the charge asymmetries ${\cal A}_{ch}( B^{+}
\ra \Nbcapkm)=0.12 \pm 0.11 \pm 0.02$ and ${\cal A}_{ch}(\Nbntoapk)=-0.16 \pm 0.12 \pm 0.01$.
The first uncertainty quoted is statistical and the second is systematic.
\end{abstract}


\maketitle
Recently the \babar\ Collaboration has reported the measurement
of the  branching fraction \cite{BRa1pi} and time-dependent \CP-violation parameters 
for the process  \Nbtoappim\  \cite{TDa1pi}. If this process were mediated by a single tree amplitude 
these measured parameters would enable a determination of the angle $\alpha$ of the unitary triangle of 
the Cabibbo-Kobayashi-Maskawa (CKM) quark-mixing matrix \cite{CKM}. However, in the presence of a penguin
amplitude with a different weak phase from the tree amplitude the measured angle, called 
 $\alpha_{\rm eff}$,  would differ from $\alpha$ \cite{CKMfit}. An upper bound on the difference
 $\Delta \alpha =| \alpha -\alpha_{\rm eff}|$ can be calculated using SU(3) together with measurements of the 
 \CP-averaged decay rates for either the decays \Nbctoapk\  or \Nbntoapk\ and the decays \NbKpi\ 
and \NbKppi\ \cite{Zupan}. Knowing the value of this difference is important in calculating bounds on the angle $\alpha$.

There are no experimental measurements of the branching fractions of the  decays \Nbctoapk\ and \Nbntoapk\ to this date.
Recent theoretical estimates of these branching fractions have been calculated assuming naive factorization 
for two different  values of the mixing angle $\theta$ between the two strange P wave axial mesons \cite{Laporta}.
The estimated branching fractions lie in the range $(16-52)\times 10^{-6}$ \cite{Laporta2}.  
Comparison between theoretical
predictions and measured quantities is useful to test the underlying
theoretical  hypotheses of factorization and $B\ra \auno$ transition form factors. 

We present measurements of the branching fraction for the decays
\Nbctoapk\ and \Nbntoapk\  with \Natopipipi\ \cite{CC}. 
We also search for a direct \CP\ violation 
by measuring the charge asymmetry ${\cal A}_{ch}$, defined as $(\Gamma^- - 
\Gamma^+)/(\Gamma^- +\Gamma^+)$, in the decay rates 
$\Gamma^{\pm}$ for a charged $B$ meson, or $\Gamma^+$ ( \Nbntoapk) and its charge conjugate
for a neutral $B$ meson.

The data were collected with the \babar\ detector~\cite{BABARNIM}
at the PEP-II asymmetric energy  \epem\ collider~\cite{pep}. An integrated
luminosity of 347~fb$^{-1}$, corresponding to
$382.9\pm 4.2$ million \BB\ pairs, was recorded at the $\Upsilon (4S)$ resonance
(on-resonance, center-of-mass energy $\sqrt{s}=10.58~\gev$).
 An additional 37~fb$^{-1}$, recorded about 40~MeV below
the $\Upsilon (4S)$ resonance (off-resonance), is used for  continuum background studies.
 
Charged particles are detected and their momenta measured by a
combination of a silicon vertex tracker, consisting of five layers
of double-sided silicon detectors, and a 40-layer central drift chamber,
both operating inside the  1.5-T magnetic field of a superconducting solenoid.
The tracking system covers 92\% of the solid angle in the center-of-mass frame.

Photons and electrons are detected  with a CsI(Tl) electromagnetic 
calorimeter.
Charged-particle identification (PID) is provided by the average
energy loss (\dedx) measured in the tracking devices and by an internally reflecting ring-imaging
Cherenkov detector (DIRC) covering the central region.
A $K/\pi$ separation of more than four standard deviations ($\sigma$)
is achieved for momenta below 3~\gevc, decreasing to 2.5 $\sigma$ at the 
highest momenta of the $B$ decay products.

Monte Carlo (MC) simulations of the signal decay modes, 
continuum, \BB\ backgrounds and detector response~\cite{geant4} are used to establish the 
event selection criteria.
Exclusive MC signal events are simulated as $B \rightarrow a_1(1260) K$
with $a_1(1260) \rightarrow \rho \pi$.
For  the  ${a_1(1260)}$ meson parameters we take the mass  
$m_0=1230$ \mevcc\ and  $\Gamma_0=400$ \mevcc\ ~\cite{evtgen,BRa1pi}.
We account for the uncertainties of these resonance parameters  in the determination of 
systematic uncertainties. 
The \Naptopipipi\ decay proceeds mainly through the intermediate states 
$(\pi \pi)_{\rho} \pi$ and $(\pi \pi)_{\sigma} \pi$ \cite{PDG2006}.
No attempt is made to separate  the contributions of the dominant P-wave 
$(\pi \pi)_{\rho}$ from the S-wave $(\pi \pi)_{\sigma}$ in the channel $\pi^+ \pi^-$. A systematic uncertainty is estimated due to the difference in the selection efficiency. 

We reconstruct the decay \Naptopipipi\ with the following requirement on the 
invariant mass: $0.87<m_{a_1}<1.5$~\gevcc\ for \Nbntoapk\ and $0.87<m_{a_1}<1.8$
~\gevcc\ for \Nbctoapk\ . The different $a_1$ mass selections are motivated by 
charm background studies. The intermediate $\pi^+ \pi^-$ state is reconstructed with 
an invariant mass between 0.51 and 1.1~\gevcc. 
Secondary  $a_1(1260)$ daughter pions  are rejected if their PID signatures
satisfy  requirements for being consistent with protons, electrons, or kaons. 
PID requirements ensure the identity of the primary charged kaon.
Candidate $\KS\to\pi^+\pi^-$ decays are formed from pairs of oppositely charged tracks with
$0.486 < m_{\pi\pi}<0.510 $ \gevcc, having a decay vertex $\chi^2$ probability 
greater than $0.001$, and a reconstructed decay length larger than three 
times  its uncertainty. 

We reconstruct the $B$-meson candidate by combining  an \auno\ candidate and a 
charged or neutral kaon. A $B$-meson candidate is characterized kinematically by the 
energy-substituted mass $\mes = \sqrt{(s/2 + \pvec_0\cdot \pvec_B)^2/E_0^2 - \pvec_B^2}$ and energy difference $\DE = E_B^*-\half\sqrt{s}$, where the subscripts $0$ and
$B$ refer to the  \UfourS\ and the $B$ candidate in the laboratory frame, 
respectively, and the asterisk denotes the \UfourS\ frame. 
The resolutions in \mes\ and  \DE\ are  about 3.0  \mevcc\ and  20 \mev\ respectively.
We require $|\DE|\le0.1$ GeV and $5.25\le\mes\le5.29\ \gevcc$. To reduce fake $B$-meson
candidates we require a $B$ vertex $\chi^2$ probability larger than  $0.01$. 
The cosine 
of the angle between the direction of the
$\pi$ meson from \NNatorhopi\  with respect to the flight direction of the $B$ 
in the \auno\ meson rest frame
is required to be between $-0.85$ and $0.85$ to suppress combinatorial
background. The distribution of this variable is flat for signal and peaks near 
$\pm 1$ for this background.

To reject continuum background, we use the angle $\theta_T$ between the thrust axis 
of the $B$ candidate and
that of the rest of the tracks and neutral clusters in the event, calculated in
the center-of-mass frame. The distribution of $\cos{\theta_T}$ is
sharply peaked near $\pm1$ for combinations drawn from jet-like $q\bar q$
pairs and is nearly uniform for the isotropic $B$-meson decays; we require
$|\cos{\theta_T}|<0.65$. The remaining continuum background is modeled from off-resonance data. 

We use MC simulations of \BzBzb\ and \BpBm\ decays to study \BB\ backgrounds, 
which can come from both charmless and charmed decays. 
The modes \Nbntoaduek and   \Nbctoaduek\  decay to the same 
final states as the signal modes. We suppress  these backgrounds with the angular 
variable \hel, defined as  the cosine
of the angle between the normal to the plane of the $3\pi$ resonance
and the flight direction of the $K$ meson     evaluated in 
the $3\pi$ resonance rest frame. Since the  $a_1(1260)$ and  $a_2(1320)$ have spins of 1 and 2 
 respectively,  the distributions of 
the variable \hel\ for these two resonances differ. We require
$|\hel|< 0.62$.

We have on average 1.3 candidates per event for  both signal decay modes
and we select the $B$ candidate with the highest $B$ vertex probability.
From the MC simulation we find that the best candidate selection algorithm
finds the correct-combination candidate in 92\,\% of both signal decay modes
and that it induces negligible bias.

We use unbinned, multivariate maximum-likelihood (ML) fits to measure
the yields of  \Nbctoapk\ and \Nbntoapk.
The likelihood function incorporates five variables.
We describe the $B$ decay kinematics with the two above-mentioned  
variables \DE\ and \mes, as well as  the invariant mass of the $3\pi$ system, a
Fisher discriminant \xf\ and the variable \hel.
The Fisher discriminant combines four
variables: the angles with respect to the beam axis in the \UfourS\
frame of the $B$ momentum and $B$ thrust axes and
the zeroth and second angular moments $L_{0,2}$ with respect to the thrust axis of the $B$-candidate \cite{Fisher}. 
Since the correlation between the ob\-ser\-va\-bles in the selected data
and in MC signal events is small, we take the probability density function 
(PDF) for each event to be a product of the PDFs for the individual 
observables. Corrections for the effects of possible correlations are made on the 
basis of MC studies described later.
The selected data samples besides the signal events contain  continuum \qq\ 
and \BB\ combinatorial background.

The \BB\ background has the 
following components in the likelihood: charmless, charm and \aduek. There are  also three
additional components:~$f_0 K$, $\rho^0 K$ 
with their yields fixed to the value determined from  the measured branching 
fractions \cite{BB}, and the non-resonant $\rho^0 \pi^+ K $ with a yield fixed in the 
fit to the value expected using an assumed branching fraction of $(2.0 \pm 2.0)\times 10^{-6}$.
We account for the uncertainties of these branching fractions in  the determination 
of the systematic uncertainties. 
A charged particle from a  signal event may be exchanged with a charged particle
 from the rest of the event. These so-called self cross feed (SCF) events 
are considered background events. The charmless \BB\ background has a dependence 
on the ML fit observables that is similar to that for SCF events, and thus the SCF 
events can be modeled as part of the charmless component. 
   
The likelihood function is defined as
\begin{eqnarray}
{\cal L} &=& \exp{\Big(-\sum_{k} n_k \Big)} \prod_{i=1}^{N} \sum_{j}  n_{j} \times \\
&&{\cal P}_k(\mes^i) {\cal P}_k(\DE^i){\cal P}_k(\xf^i){\cal P}_k(m_{a_1}^i){\cal P}_k({\hel}^i)\,, \nonumber
\label{eqn:likelihoodtag}
\end{eqnarray}
where $N$ is the total number of events in the fit sample, 
$n_k$ is the yield fitted for the likelihood component $k$ and $P_k(x^i)$ is the PDF 
for observable $x$ in event $i$ . We determine the PDFs 
for signal and \BB\ backgrounds from MC distributions in each observable. For the 
continuum background we establish the functional forms and initial parameter values of 
the PDFs with off-resonance data. The PDF of the invariant mass of the ${a_1(1260)}$ 
meson in  signal events is parameterized as a relativistic Breit-Wigner
lineshape with a mass-dependent width which takes into account the effect of the 
mass-dependent $\rho$ width~\cite{WA76}. We fix the \auno\ meson
parameters 
to the values found in the branching fraction measurement 
of $B^0\rightarrow a_1^{\pm} \pi^{\mp}$\cite{BRa1pi}.  
The PDF of the invariant mass of the ${a_2(1320)}$ meson is parameterized by a 
relativistic Breit-Wigner distribution. The \mes\ and \DE\ distributions for  signal are 
parameterized as double Gaussian functions. The \DE\ distribution for continuum 
background is parameterized by a linear function. The \mes\
distribution for the combinatorial background is described by an empirical function
that accounts for threshold effects~\cite{argus}. 
We model the Fisher distribution \xf\ using a Gaussian function with different 
widths above and below the mean. The \hel\ distributions are modeled using 
polynomials. 

In the fit for the decay \Nbntoapk\  (\Nbctoapk) there are respectively fourteen (twelve) free 
parameters: five (five) yields and nine (seven) parameters affecting the shape 
of the combinatorial background. Table I lists the results of the fits. We measure the 
signal yield bias by generating and fitting MC
simulated samples containing signal and background populations
expected from data. The signal reconstruction  efficiency is obtained
from the  fraction of correctly reconstructed signal MC events passing the selection criteria.
 Branching fractions for each decay are computed by
subtracting the fit bias from the measured yield, and dividing the
result by the efficiency,  the daughter branching fraction
  product, and the number of \BB\ pairs produced.  Equal
production rates to \BzBzb\ and \BpBm\ pairs are assumed. The
significance is taken as the square root  of the difference between
the value of  $-2\ln{\cal L}$ (with systematic uncertainties included)
for zero signal and the value at its minimum.
 
\begin{table}[!htb]
\label{tab:results}
\caption{Number of events $N$ in the sample, fitted signal yield and
  measured bias (to be subtracted from the signal yield) in events (ev.),
  detection  efficiency ($\epsilon$), daughter branching fraction
  product $\prod\calB_i$, significance ($S$) (systematic
  uncertainties included),     the products of the branching fractions
\BraknPiPiPi\  and \BrakcPiPiPi\ respectively,
and charge asymmetry 
  with statistical and systematic error.}
\begin{center}
\begin{tabular}{lcc}
\hline
Parameter               & \Nbnapkmm\         & \Nbcapkm\ \\
\hline
N (ev.)                 &$12196$            & $9468$ \\
Signal yield (ev.)      &$272 \pm 44$   & $241 \pm 32$ \\
Bias (ev.)              &$+24$          & $+18$ \\
$\epsilon$ (\%)         &$7.9 $             & $9.6$ \\
$\prod\calB_i$ (\%)     &$100.0$            & $34.6$\\
$S$($\sigma$)           & $5.1$             & $6.2$\\
${\cal B}(\times 10^{-6})$ &$8.2 \pm 1.5 \pm 1.2 $ &$ 17.4 \pm 2.5 \pm 2.2$ \\
${\cal A}_{ch}$        &  $-0.16 \pm 0.12 \pm 0.01$    & $0.12 \pm 0.11 \pm 0.02$     \\
\hline
\end{tabular}
\end{center}
\end{table}

\begin{figure}[!h]
\resizebox{\columnwidth}{!}{
\includegraphics[]{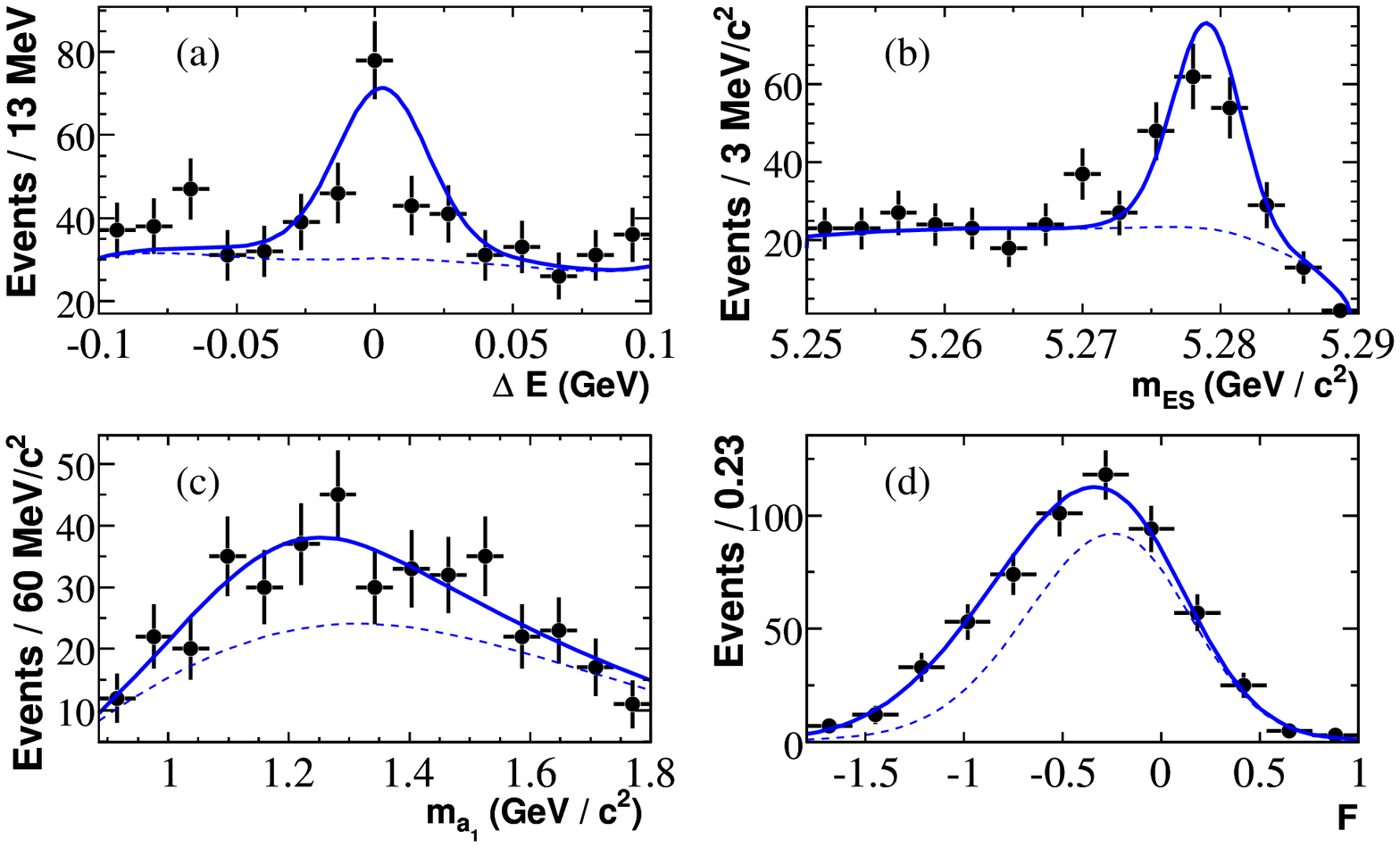} 
}
\caption{Projections of a) \DE, b)  \mes, c) $m_{a_1}$, and d) \xf\ for the \Nbctoapk\ decay mode. Points represent on-resonance data, dashed lines 
the continuum and \BB\ backgrounds, and solid lines the full fit
function. These plots are made with a cut on the signal likelihood which includes $30\%-40\%$ of the signal.}.

\label{fig:projections}
\end{figure}

\begin{figure}[!h]
\resizebox{\columnwidth}{!}{
\includegraphics[]{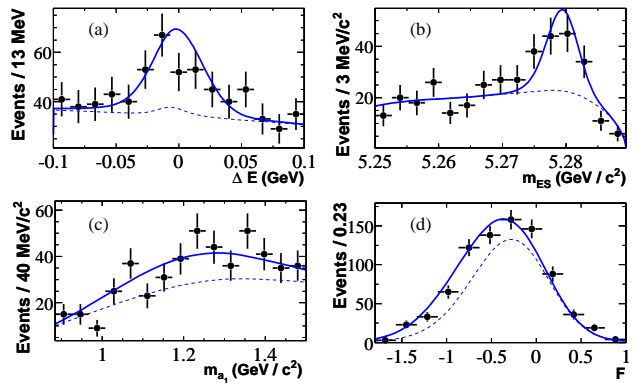} 
}
\caption{Projections of a) \DE, b)  \mes, c) $m_{a_1}$, and d) \xf\ for the \Nbntoapk\ decay mode. 
Points represent data, dashed lines 
the continuum and \BB\ backgrounds, and solid lines the full fit
function. These plots are made with a cut on the signal likelihood
which includes $30\%-40\%$ of the signal.}.

\label{fig:projections2}
\end{figure} 

In Figs.~\ref{fig:projections} and~\ref{fig:projections2} we show the \DE, \mes, $m_{a_1}$, and \xf\  
projections made by selecting events with a signal likelihood (computed without the variable
shown in the figure) exceeding a threshold that optimizes the
expected sensitivity.

Most of the systematic errors on the signal yield arising from  
uncertainties in the values of the PDF parameters are already incorporated 
into the overall statistical error, since they are floated in the fit.
The systematic error on the \Nbntoapk\ (\Nbctoapk) fit yield 
is $28$ ($23$) events, which is obtained by varying the PDF parameters within 
their uncertainties. We estimate the uncertainty arising from  the number of \BB\ pairs 
to be $1.1\%$. The uncertainty in the fit bias correction is $12$ ($9$) events, 
taken as
half of the fit bias correction. The uncertainty in the \auno\  meson parameters 
is $12$ ($6$) events. 
The systematic uncertainty
assigned to the fixed yields in the fit is $3$ ($4$) events. The systematic effect due to 
differences between data and MC for the \costhr\ selection is
$1.8\%$. A systematic uncertainty of $2.0\, (2.5)  \,\%$  is
evaluated for the PID. The tracking efficiency contributes to the
systematics with  $1.8\, (1.3)  \,\%$. 
A systematic uncertainty of  $2.5\,\%$ is estimated for the difference in 
selection efficiency in the decay modes through the dominant P-wave 
$(\pi \pi)_{\rho}$ and the S-wave $(\pi\pi)_{\sigma}$. The contribution of 
interference between $a_2(1320)$ and $a_1(1260)$ is negligible. In fact, varying 
the $a_2(1320) K$ background with different selection criteria on the angular 
variable \hel\ gives no significant change to the efficiency-corrected 
signal yield of $a_1(1260) K$. We find also that the systematic effect due to 
different form factors in MC signal simulation is negligible. The
total systematic  error on the branching fraction of the charged (neutral) mode  is $14\%$ ($13\%$).

The primary sources of systematic  uncertainties  in the charge asymmetry
measurement are the track reconstruction or particle identification,  the
imperfect modelling of the interactions with material in the
detector and the \BB\ background. We study these systematic
uncertainties with MC signal events, \qq\ background in the data, and
control samples. We assign a systematic uncertainty of $0.02$ ($0.01$)
to the charge asymmetry of the charged (neutral) mode.

In summary, we have measured the branching fractions 
$\calB(\Nbctoapk)$$\calB(\Naptopipipi)$ = \Rakc\ and $\calB(\Nbntoapk)$ $\calB(\Namtopipipi)$ = \Rakn . 
The charged (neutral) $B$ decay mode
 is observed with a significance of 6.2 (5.1) standard deviations, which
 includes systematic uncertainties. We find no evidence for a direct 
\CP-violating asymmetry  in these  decay modes.  
Assuming ${\cal B}(a^{\pm}_1(1260) \rightarrow
\pi^{\mp} \pi^{\pm} \pi^{\pm})$ is equal to ${\cal B}(a^{\pm}_1(1260)
\rightarrow \pi^{\pm} \pi^0 \pi^0)$, 
and that  ${\cal B}(a^{\pm}_1(1260)\rightarrow (3\pi)^{\pm})$ is
equal to 100\%~\cite{PDG2006}, we obtain  ${\cal B}(B^0 \rightarrow
a^{-}_1(1260) K^{+})=(16.3 \pm 2.9 \pm 2.3)\times 10^{-6}$ 
and ${\cal B}(B^{+} \rightarrow a^{+}_1(1260) K^{0}=(34.9 \pm 5.0 \pm 4.4)\times 10^{-6}$.
These results are in reasonable agreement with the theoretical estimates.

\input acknow_PRL

\end{document}

%% file: authors_jul2007.tex
%
\author{B.~Aubert}
\author{M.~Bona}
\author{D.~Boutigny}
\author{Y.~Karyotakis}
\author{J.~P.~Lees}
\author{V.~Poireau}
\author{X.~Prudent}
\author{V.~Tisserand}
\author{A.~Zghiche}
\affiliation{Laboratoire de Physique des Particules, IN2P3/CNRS et Universit\'e de Savoie, F-74941 Annecy-Le-Vieux, France }
\author{J.~Garra~Tico}
\author{E.~Grauges}
\affiliation{Universitat de Barcelona, Facultat de Fisica, Departament ECM, E-08028 Barcelona, Spain }
% ...
\collaboration{The \babar\ Collaboration}
\noaffiliation

%% file: acknow_PRL.tex
 We are grateful for the excellent luminosity and machine conditions
provided by our \pep2\ colleagues, 
and for the substantial dedicated effort from
the computing organizations that support \babar.
The collaborating institutions wish to thank 
SLAC for its support and kind hospitality. 
This work is supported by
DOE
and NSF (USA),
NSERC (Canada),
IHEP (China),
CEA and
CNRS-IN2P3
(France),
BMBF and DFG
(Germany),
INFN (Italy),
FOM (The Netherlands),
NFR (Norway),
MIST (Russia), and
STFC (United Kingdom). 
Individuals have received support from CONACyT (Mexico), 
Marie Curie EIF (European Union),
the A.~P.~Sloan Foundation, 
the Research Corporation,
and the Alexander von Humboldt Foundation.

%% file: paper.bbl
\begin{thebibliography}{99}
\bibitem{Laporta2}
For the decay  \Nbctoapk\  the mentioned estimated range of branching 
fractions is obtained using  the experimental measurement of the branching 
fraction of the channel $B^+ \ra  \pi^+ K^0$. See Ref.~\cite{Laporta} 
for more details.
\bibitem{CC}
Charge-conjugate reactions are implied throughout this paper.
\end{thebibliography}
